\documentstyle[12pt,epsfig]{article}
\def\issue(#1,#2,#3){#1 (#3) #2} 

\def\opcit(#1){ {\em op. cit.}, #1}

\def\APP(#1,#2,#3){Acta Phys.\ Polon.\ \issue(#1,#2,#3)}
\def\ARNPS(#1,#2,#3){Ann.\ Rev.\ Nucl.\ Part.\ Sci.\ \issue(#1,#2,#3)}
\def\CPC(#1,#2,#3){Comp.\ Phys.\ Comm.\ \issue(#1,#2,#3)}
\def\CIP(#1,#2,#3){Comput.\ Phys.\ \issue(#1,#2,#3)}
\def\EPJC(#1,#2,#3){Eur.\ Phys.\ J.\ C\ \issue(#1,#2,#3)}
\def\EPJD(#1,#2,#3){Eur.\ Phys.\ J. Direct\ C\ \issue(#1,#2,#3)}
\def\IEEETNS(#1,#2,#3){IEEE Trans.\ Nucl.\ Sci.\ \issue(#1,#2,#3)}
\def\IJMP(#1,#2,#3){Int.\ J.\ Mod.\ Phys. \issue(#1,#2,#3)}
\def\MPL(#1,#2,#3){Mod.\ Phys.\ Lett.\ \issue(#1,#2,#3)}
\def\NP(#1,#2,#3){Nucl.\ Phys.\ \issue(#1,#2,#3)}
\def\NIM(#1,#2,#3){Nucl.\ Instrum.\ Meth.\ \issue(#1,#2,#3)}
\def\PL(#1,#2,#3){Phys.\ Lett.\ \issue(#1,#2,#3)}
\def\PRD(#1,#2,#3){Phys.\ Rev.\ D \issue(#1,#2,#3)}
\def\PRL(#1,#2,#3){Phys.\ Rev.\ Lett.\ \issue(#1,#2,#3)}
\def\SJNP(#1,#2,#3){Sov.\ J. Nucl.\ Phys.\ \issue(#1,#2,#3)}
\def\ZPC(#1,#2,#3){Zeit.\ Phys.\ C \issue(#1,#2,#3)}

\newcommand{\ml}{m_{\tilde{\ell}_{L}}}
\newcommand{\mr}{m_{\tilde{\ell}_{R}}}

\newcommand{\ee}{e^+e^-}

\newcommand{\sw}{x^l_W }

\def \snu{\tilde\nu}  
\def\ET {\not\!\!{E_T}}
\def\wino{\tilde \chi^{\pm}}   
\def\Chl{\tilde \chi^{0}_1}   
\def\Chh{\tilde \chi^{0}_2}   
\def\Lsp1{\tilde \chi^{0}}   
\begin{document}
\baselineskip=17pt
\begin{flushright}
PM/02--28\\
October 2002\\
\end{flushright}
\vspace*{0.9cm}

\begin{center}

{\large\bf Are light sneutrinos buried in LEP data?}

\vspace{0.7cm}
Amitava {\sc Datta}\footnote{Email: adatta@juphys.ernet.in} \\
\vspace{0.1cm}
Department of Physics, Jadavpur University, Kolkata 700032, India

\vspace{0.3cm}
and
\vspace{0.3cm}

Aseshkrishna {\sc Datta}\footnote{Email: datta@lpm.univ-montp2.fr} \\
\vspace{0.1cm}
Laboratoire de Physique Math\'ematique et Th\'eorique, UMR5825--CNRS,\\
Universit\'e de Montpellier II, F--34095 Montpellier Cedex 5, France

\end{center}

\vspace*{1cm}

\begin{abstract}
\noindent
Supersymmetry may resolve the disagreement between the
precision electroweak data and the direct limit on the
higgs mass,
if there are  light sneutrinos in the mass range
55 GeV $ < m_{\snu} < $ 80 GeV. Such sneutrinos
should  decay invisibly  with 100\% branching ratio and contribute to the
$\gamma$ + missing energy signal,  investigated  by all the LEP
groups. It is shown that while the data accumulated by a single group
may not be adequate to reveal such sneutrinos, a combined analysis
of the data collected by all  four groups will be  sensitive to
$m_{\snu}$ in the above range.  If no signal
is found a   lower  bound  on $m_{\snu}$
stronger than that obtained from the $Z$-pole data may emerge.
\end{abstract}
\newpage

\section{Introduction}
It is generally believed that  the precision electroweak(EW) data are in good agreement 
with  the  standard model(SM) of particle physics\cite{ewp}. Yet one has to 
admit that during the last couple of years a subset of  the data has made the
situation somewhat uncomfortable for the SM\cite{chano1,alta,chano2}. 
If only a small  subset of a large volume of data is in strong disagreement with theory, 
such that its removal improves the quality of the fit, then 
one would normally believe that either statistical fluctuations
or some hitherto unknown systematic errors  might have  affected  this subset. 
A conservative approach  would then be to 
keep the subset in the cold storage awaiting improved statistics or a better 
understanding of the systematics, rather than invoking a new physics model.
   
The situation, unfortunately, is not that simple in the present case. It is now well known 
that the effective leptonic weak mixing angle ( given by
$sin^2 \theta^l_W$ = $x^l_W$) determined
from the leptonic asymmetries measured by the  LEP groups and SLD are in severe 
disagreement with the same parameter as determined by the hadronic asymmetries\cite{ewp}.
The discrepancy in  the global analysis becomes more prominent  if the effective on-shell weak mixing 
angle extracted from recent  neutrino scattering data\cite{nutev} is taken into account\cite{chano2}. 
In Ref.\cite{ewp} several standard model fits were performed (see Table 13.2 ) 
 and it was noted
that the $\chi^2$/d.o.f was large in each case. The large dispersion in the fitted value
of $\sw$ from various asymmetries, responsible for the poor fit, was interpreted as the 
result of 
fluctuations in one or more of the experimental inputs and 
the issue was not pursued any further. Unfortunately  no improvement in statistics is 
expected in the
near future. Most of the analyses based on the measurements at the $Z$-pole are now more or less final\cite{ewp}.
With more than 700 pb$^{-1}$ of data per experiment
at energies higher than the $Z$-pole  an improved accuracy in  
$m_W$ measurement is expected which in turn may lead to a better understanding of the SM
vis a vis LEP data.

 If the hadronic asymmetries are  discarded the quality of the fit improves dramatically, as 
expected, but the
mass of the Higgs boson obtained from the fit turns out to be much smaller\cite{chano1,alta,chano2}
than the lower limit obtained from the direct searches  at LEP\cite{higgs}. 
Given the theoretical
uncertainties (unknown  higher order corrections, the precise value of 
an important input -  the fine structure constant 
evaluated at $m_Z$ etc.) nicely reviewed in Ref.\cite{ewp}, the possibility of statistical fluctuations 
and poorly understood systematics, it is not impossible  that the data can still be in agreement with
the SM. However, as analyzed in great details in Ref.\cite{chano2}, something totally unexpected has to
happen. For example, if statistical fluctuation is the possible explanation then it is imperative 
that not only the measurements in disagreement ( some of the hadronic asymmetries, say ) 
with the SM but also the ones which have been thought
to be the evidence for the SM for so many years must involve large fluctuations. It was, therefore, 
argued in Ref.\cite{chano2} that new physics seems to be the favoured solution, although  the evidence is
not fully conclusive.

Several new physics models have already been proposed as possible solutions of the alleged conflict
between the SM and the data
(see  Ref.\cite{chano2} for a list of  further references). In this paper the focus will be on  
the supersymmetric solution\cite{alta}. This solution seems to be attractive because
 one does not need supersymmetry only to ameliorate the malady
in the precision electroweak data. It is needed to answer deeper issues. It solves
the hierarchy puzzle that haunts non-supersymmetric grand unified theories and facilitates the 
coupling constant unification. The supersymmetric extension of the minimal standard model
(MSSM), therefore, seems to be a well motivated step beyond the SM. 
It was shown in Ref.\cite{alta} that the quality of the fit to the precision electroweak data
as well as the agreement with the lower  bound on the Higgs mass improve in the MSSM with light
slepton-sneutrinos. 
In particular, the MSSM with sneutrino ($\snu$)s having  mass in the range $55-80$ GeV
seems to be preferred by the data\cite{alta}.The  left
slepton  masses ($\ml$) are related to $m_{\snu}$ by the SU(2) breaking D-term in a model independent way.
 In order to make them
heavier than the LEP bound moderate or large $\tan\beta$ was needed 
The fit favoured much heavier squarks (mass $\sim$
1 TeV), was practiclly independent of $m_A$ (the mass of the pseudoscalar Higgs boson) and $\mr$
(the mass of the right slepton). Gaugino mass unification was assumed although the U(1) gaugino
mass parameter had little impact. Mass of the lighter chargino ($m_{\wino}) < $ 
 150 GeV  improved the fit
although higher masses also had  reasonable agreement with the data. Left sleptons and sneutrinos
belonging to different generations were assumed to be degenerate. .  

The emphasis of this paper would be on  the possibility of testing
 the light sneutrino  hypothesis via  direct searches using existing data. 
This is especially important since any immediate
 improvement in the indirect test of this scenario using electroweak data is not feasible.
It is gratifying to note that the existing LEP data on single $\gamma$ + missing energy events
\cite{photon} can indeed shed light on this issue. 

However, before taking up the main issue, we want to review  briefly
a related topic. Is there  
 a theoretically well motivated  supersymmetry breaking mechanism
which  leads naturally  to the  light slepton scenario?
 It was noted in Ref.\cite{alta} that this  scenario cannot be 
accommodated in the popular minimal supergravity (mSUGRA) model 
where supersymmetry breaking is 
driven by gravity mediated interactions leading to a common scalar mass ($m_0$) 
at a scale  which is often assumed, somewhat arbitrarily,  to be the GUT scale ($M_G$). 
In such models the masses  of the  sleptons and the sneutrinos are correlated. As a result
$m_{\snu}$ in the above mass range would inevitably
lead to a somewhat lighter right slepton with  mass ($\mr$) in conflict with the 
existing  LEP lower bound.

This observation has recently been quantified by the ALEPH collaboration. Their
work is based on a model, similar to mSUGRA,  with a common $m_0$ and a common gaugino
 mass at the GUT  scale\cite{aleph}. The limit is based on  
the  data for various sfermion-gaugino searches. No direct  sneutrino signal 
was searched for. Yet from the absence of any slepton signal they 
 obtained a model dependent  lower bound on the sneutrino mass ($m_{\snu} >$ 84 GeV) 
by exploiting the correlation between  slepton and sneutrino masses.
 
Possible alternative scenarios 
 with light sleptons  were qualitatively discussed  in Ref.\cite{alta}. It was noted that 
if $m_0$ is generated at
the Planck scale ($M_P$) instead of $M_G$ then the running between $M_P$ and $M_G$ may 
indeed lead to a somewhat larger $\mr$ at the weak scale. This  
avoids  the conflict with the LEP bound. In fact an inspection  of this running 
within the framework of an SU(5) SUSY GUT, as given
in Ref.\cite{poma}, would encourage this scenario. The other viable model proposed was the 
anomaly mediated  supersymmetry breaking (AMSB)  model\cite{amsb}. 
In the latter  model, without any additional assumption, the slepton-sneutrino 
masses turn out  to be tachyonic. In the simplest version of this model one
 adds a  common soft breaking  term  making the mass squared terms positive. 
The slepton mass can, therefore, be 
arbitrarily small and apparently
the light slepton-sneutrino scenario can be accommodated.

Both the above solutions, however, lead to an unstable  electroweak symmetry 
breaking vacuum\cite{abhi1,abhi2} as the potential becomes unbounded from 
below\cite{casas}. In fact in
the case of the AMSB model the requirement of vacuum stability leads to a lower bound on the
slepton mass which after allowing for all theoretical uncertainties is approximately 300 GeV\cite{abhi2}. Of course one can argue that we are living in a false vacuum with a life time larger 
than the age of the universe\cite{false1} which makes the requirement of vacuum stability  
redundant. However, it is difficult to accept
this solution uncritically. First of all the calculation of the  probability of tunnelling
from the false vacuum to the true one, which
is rather straightforward in a model with a single scalar, becomes far too complicated in 
the MSSM with multiple scalars. In fact the earlier calculations of this probability have
been criticized by  more recent ones (see the second paper of Ref.\cite{false1}). Yet there is no 
way of testing   the reliability of the recent calculations 
 as they cannot be verified  experimentally. Furthermore tunnelling being a
probabilistic phenomenon, the unpleasant possibility that charge and colour symmetry will be broken
at the very next moment always remains open. In our opinion, therefore, the false vacuum scenario 
should remain as a theoretical curiosity unless the complete determination of the sparticle spectrum 
in  future experiments points  unmistakably to a set of SUSY parameters leading to
an unstable electroweak symmetry breaking vacuum.  

In view of the above discussions it is prudent to look for a model with light sleptons
without jeopardizing the stability of the vacuum. Such a scenario arises when an SO(10)
SUSY GUT  directly breaks down to the SM gauge group\cite{abhi1}. Certain U(1) 
symmetry breaking $D$-terms at $M_G$\cite{drees} can then lead  to a sparticle spectrum with right sleptons naturally
heavier than the left slepton/sneutrinos. In fact it was shown in Ref.\cite{abhi1} that in this 
model one may have sneutrinos in the mass range preferred by the  precision electroweak data 
while the right selectron mass is beyond  the kinematic reach of LEP. 

This is not to suggest that the above model is the only  or even the most
appealing model  with light sneutrinos. It simply demonstrates that the physics at $M_G$
or $M_P$ involves too many uncertainties 
to determine precisely from the boundary conditions at $M_G$, 
what the sparticle spectrum at low energies should look like.
 Any mass spectrum  apparently preferred by the data  should, therefore, be taken 
seriously irrespective
of being favoured or disfavoured by the currently prevailing theoretical prejudices. However,
further direct experimental tests to confirm the spectrum is urgently needed. We shall 
now turn our attention to this task.

If the sneutrino and other sparticle masses are 
indeed in the range preferred by the precision electroweak data
then the $\snu$ would  be the next lightest supersymmetric particle (NLSP), the lightest
neutralino ($\Chl$) being the lightest supersymmetric particle (LSP). Such a sneutrino will decay into 
the
invisible mode $\snu \rightarrow \nu \Chl$ with 100\% branching ratio (BR). As such sneutrinos 
 will act as carriers of missing energy just like  the $\Chl$ in a 
$R$-parity conserving model, they have been 
called virtual lightest supersymmetric particle (VLSP)  or the effective lightest
supersymmetric particle (ELSP)\cite{vlsp}. Such sneutrino pairs produced in association with
an energetic photon will lead to the signal $\ee \rightarrow \gamma + $ missing energy\cite{asesh1}
This signal
if seen over the SM background may indicate the existence of light sneutrinos. In the MSSM
there are,  however, other processes leading to the same final state. Of course the
processes $\ee \rightarrow \gamma \Chl \Chl $ is present in all versions of the $R$-parity
conserving model. If the second lightest neutralino ($\Chh$) decays into the channel
$\Chh \rightarrow \snu  \nu$ with a large BR, then all processes belonging
to the class $\ee \rightarrow \gamma \Lsp1_i \Lsp1_j $ (i = 1,2), will 
also contribute to the signal. The actual value of this BR, however, is more model dependent
as it depends also on $\mr$.
       
The cross sections of all supersymmetric contributions to  $\ee \rightarrow \gamma + $ missing 
energy were exactly computed in Ref.\cite{asesh1}. It was shown  that with special 
kinematic cuts (see below)  the signal can be seen over the SM background. However, the
results of\cite{asesh1} were computed on the basis of 500 pb$^{-1}$ of integrated
luminosity collected at $\sqrt{s} = 190$ GeV. More recently LEP has run over a more varied
range of $\sqrt{s}$ including energies considerably higher than  $\sqrt{s} = 190$ GeV.
The total accumulated luminosity\cite{lumi}  also happens to be much larger than that 
considered in Ref.\cite{asesh1}. Since the  sneutrino VLSP signal  
assumes new significance in the light of the precision
electroweak data, a re-examination of the analysis of  Ref.\cite{asesh1} is called for.

It may be recalled that all the  LEP groups have extensively studied 
the single $\gamma$ + missing energy signal\cite{photon}. 
In most of the analyses  the cross section of the SM process $\ee \rightarrow  \nu 
\bar{\nu} \gamma$ was measured and the result was then used for neutrino counting. The kinematical
cuts were  optimized to suppress the backgrounds to this  process (radiative
Bhabha scattering etc.). This  process 
receives a large contribution when the photon energy is such that an on shell $Z$ is 
produced which subsequently decays to a $\nu \bar{\nu}$ pair. An appropriate upper cut
on the photon energy excludes  this radiative return to the $Z$-pole and drastically
reduces this cross section\cite{asesh1}. 
However, in the presence of initial state radiations the 
photon energy appropriate for radiative return is somewhat smeared and the efficiency of the 
cut is  reduced to some extent (see below). Nevertheless this cut is very useful to improve the 
signal to background ratio. In a limited number of new physics searches such cuts were also
employed by some of the LEP collaborations. For example DELPHI\cite{delphi1} looked for the production
of a pair of superlight gravitinos in association with a photon, where $E_\gamma$ was restricted
to $E_\gamma <$ 50 GeV. 
The dynamics and kinematics of this process  is, however, quite different  
from the production of a  pair of heavy sneutrinos and their conclusions  cannot be extended 
to the case under study in a straightforward way.

In the following we shall repeat the analyses of Ref.\cite{asesh1} for realistic energies and
luminosities.  We,  however, find that
the data collected at a particular energy or by a particular group is not enough to  produce
a signal to background ratio which is statistically significant. On the other hand, if the data
from all the groups at different energies are combined  more significant  results may be obtained.
Admittedly this is a difficult task and should be carried out by the experts. In particular 
combining the systematics of different experiments requires special care.
 Moreover, the integrated luminosities accumulated by each group at a given energy
are not always available.  For example, in the energy range $\sqrt{s}$ = 191.6 - 201.6 GeV
the total luminosity collected by OPAL is said to be 212 pb$^{-1}$. We could not find
out further break-ups. 
On the other hand the break-ups of the total luminosity collected by the other groups 
at different energies in this range were available in most cases 
(see Table 1)\cite{lumi}.

The purpose of our simple minded analysis is, therefore, not to obtain rigorous bounds. 
We are rather interested in  illustrating  the sensitivity of the existing data
to  sneutrino masses in the range preferred by the precision electroweak data. 
We, therefore, often take recourse to various approximations.
For example,  we roughly estimate the OPAL 
luminosity at a particular $\sqrt{s}$, by scaling the corresponding quantity 
given by  another group,
by the ratio of the total luminosity collected by OPAL and the other group in the entire energy range.
Whenever the detailed break-ups were not available, we employed  this approximation. The information
 that we could gather from the literature\cite{lumi} is summarized in Table 1:       


\begin{center}
\renewcommand{\arraystretch}{1.5}
\begin{tabular}{|c|c|c|c|c|c|}
\hline
$\sqrt{s}$(GeV) & ALEPH & DELPHI & L3 & OPAL & Total $\int{\cal L}$ (pb$^{-1}$) \\
\hline
\hline
188.6 & 173.6 & 154.7 & 176.4 & 177.3 & 682.0 \\
\hline
191.6 & 28.9 & 25.9 & 29.5 & $\sim$ 25.7 & $\sim$ 110.0 \\
\hline
195.5 & 79.9 & 76.4 & 83.0 & $\sim$ 73.0 & $\sim$ 312.3 \\
\hline
199.5 & 87.0 & 83.4 & 82.1 & $\sim$ 76.8 & $\sim$ 329.3 \\
\hline
201.6 & 44.4 & 40.6 & 36.9 & $\sim$ 36.4 & $\sim$ 158.3 \\
\hline
203.7 & $\sim$ 6.9 & 8.4 & $\sim$ 8.5 & $\sim$ 6.3 & $\sim$ 30.1 \\
\hline
205.2 & 75.3 & 76.2 & $\sim$ 77.9 & $\sim$ 57.8 & $\sim$ 287.2 \\
\hline
206.7 & 122.6 & 121.6 & $\sim$ 125.6 & $\sim$ 93.2 & $\sim$  463.0 \\
\hline
208.2 & 9.4 & 8.3 & $\sim$ 9.1 & $\sim$ 6.8 & $\sim$  33.6 \\
\hline
\hline
Total & 628.0 & 595.5 & $\sim$ 629.0 & $\sim$ 553.3 & 2405.8 \\
\hline
\end{tabular}
\end{center}
Table 1. The breakups for luminosities accumulated by the LEP groups in different
$\sqrt{s}$-bins in the range $188.6 < \sqrt{s} < 208.2$ GeV. The approximated 
luminosities are indicated explicitly and follow from the treatment explained in
the text.
\vskip 5pt
In Table 2 we present a sample analysis. For this analysis we assume  gaugino mass unification
and the slepton masses to be flavour independent.
The  SUSY parameters are chosen to be $m_{\snu}$  = 56 GeV, $M_2$ (the SU(2) gaugino mass) = 110 
GeV,
$\mu$ (the higgsino mass parameter) = $-300$ GeV and $\tan\beta$ (the ratio of the vacuum expectation
values of the two Higgs bosons) = 10. These immediately lead to the following mass pattern:
 $\ml$
(the mass of the left slepton) = 96.96 GeV, $m_{\wino}$ 
(the mass of the lighter chargino) = 106.5 GeV, $m_{\Chl}$ (mass of the lightest
neutralino) = 55.1 GeV and $m_{\Chh}$ (the mass of the second lightest neutralino) = 106.3 GeV.
The mass of the right slepton $(\mr)$ cannot be fixed without further assumption. We 
assume it to be 100 GeV. As discussed above  $\Chh$ can  decay into charged leptons and sleptons 
and, hence, may
 not decay into the invisible channel $\nu \snu$  with 100\% BR. In order to obtain a conservative
estimate we, therefore, ignore the  $\Chl + \Chh$ + $\gamma$ events,  although a 
significant fraction of such events may also lead to the signal under study. Thus in practice the 
signal in Table 2 can be somewhat larger than what has been  shown.  

The kinematic cuts imposed are as follows. Only hard photons emitted into the angular interval
 $15^\circ < \theta_\gamma < 165^\circ$ are considered, where $\theta_\gamma$ is the polar angle
of the photon with respect to the beam direction. Roughly speaking this 
 covers   the barrel 
 and the end cap regions of a typical LEP detector. 
Of course, the detailed geometries are different 
for different experiments. The detection efficiency of photons, which is detector dependent,
is taken to be unity for simplicity. There is  a lower cut on the photon energy
$E_\gamma$ (or $p_{T_\gamma}$) which along with the above angular cuts reduces the background
from radiative Bhabha scattering and other processes.
       This  lower cut on photon energy is  parametrized
      by a variable $x_\gamma$ defined to be $x_\gamma=E_{\gamma}/E_{beam}$. Typically
      the LEP experiments require $x_\gamma > 0.06$ in the barrel regions  and
      $x_\gamma > 0.1$ in the end cap  regions. As we are trying to include the
      end cap region in our analysis we conservatively set $x_\gamma > 0.09$ for all  
      $\theta_\gamma$  which is likely to compensate for the gaps 
       actually present  between the 
      end cap and  the barrel regions that we neglect in our analysis. 
      In addition there should be an upper cut on the 
      photon energy to exclude events with 
       radiative return to the $Z$-pole. Again, in principle, this
      cut should be a function of $\sqrt{s}$ and $m_{\snu}$ to be probed. But in our simple minded analysis 
      the photon energy
      is restricted to the range  $E_{min} < E_\gamma < 60$ GeV at all c.m. energies. The signal 
      to background ratio can be further improved by optimizing  the upper cut with beam energy
      and $m_{\snu}$. 
   
\vskip 5pt
\begin{center}
\renewcommand{\arraystretch}{1.6}
\begin{tabular}{||c|c|c|c|c|c|c||}
\hline
\hline
 $\sqrt{s}$ & $\sigma_{\nu \bar{\nu} \gamma}$  & 
$\sigma_{\tilde{\nu} \tilde{\nu}^* \gamma}$ & 
$\sigma_{\chi_1^0 \chi_1^0 \gamma}$ & $\int {\cal L}$ & Number & Number  \\
(GeV) & (pb) & (pb) & (pb) & (pb$^{-1}$) & of BG Events & of Signal Events \\
\hline
\hline
188.6 & 1.303(2.256) & 0.101 & 0.018 & 682 & 886(1534) & 81 \\
\hline
191.6 & 1.268(2.162) & 0.103 & 0.019 & 110 & 139(238) & 13 \\
\hline
195.5 & 1.251(2.040) & 0.105 & 0.019 & 312 & 390(636) & 39 \\
\hline
199.5 & 1.222(1.949) & 0.106 & 0.020 & 329 & 402(641) & 41 \\
\hline
201.6 & 1.208(1.894) & 0.107 & 0.020 & 158 & 188(295) & 20 \\
\hline
203.7 & 1.201(1.851) & 0.107 & 0.020 & 30 & 36(56) & 4 \\
\hline
205.2 & 1.193(1.828) &  0.108 & 0.020 & 287 & 338(517) & 36 \\
\hline
206.7 & 1.188(1.794) & 0.108 & 0.021 & 463 & 546(825) & 59 \\
\hline
208.2 & 1.179(1.771) & 0.108 & 0.021 & 33 & 38(57) & 4 \\
\hline
 Total & & & & & 2963(4599) & 297 \\
\hline
\hline
\end{tabular}
\end{center}
\vskip 10pt
Table 2.  The signal and the background rates for different values of $\sqrt{s}$.  
The kinematic cuts employed  are $15^\circ < \theta_\gamma < 165^\circ$,
$E_\gamma^{min} < E_\gamma < 60$ GeV where $E_\gamma^{min}=x_\gamma E_b$. 
The SUSY parameters are $\mu= -300$ GeV,
$\tan\beta = 10$ and $M_2=2 M_1 = 110$ GeV, $m_{\tilde{\nu}}=56$ GeV ,  
$m_{{\tilde e}_L} = 96.96$ GeV and $m_{{\tilde e}_R}= 100$ GeV.

\vskip 5pt
A few remarks  on the method of calculation are now in order. It was noted in Ref.\cite{asesh1}
that the cross section for $\ee \rightarrow \nu \bar{\nu} \gamma$ 
calculated by us was smaller than the cross 
sections reported by other groups. Subsequently we found a sign  error in one of the 
interfering amplitudes and  our results are now in exact agreement with the other groups. 
The cross section corrected for initial state radiation as discussed in Ref.\cite{asesh1},
agrees nicely with  the result obtained by using the 
package {\tt CompHEP}\cite{comp}. There was also a sign error in one 
of the interfering amplitudes for the process $\ee \to \snu \snu^* \gamma$ which  
underestimated  the signal cross section. Since both the signal and the background
 were underestimated in  Ref.\cite{asesh1}, the conclusions were roughly correct albeit somewhat
fortuitously. After corrections our result for  $\ee \to \snu \snu^* \gamma$  agrees
with that from {\tt CompHEP}. 
Also, the cross sections of $\Chl$  pair + $\gamma$ production   
is  in agreement with results from {\tt CompHEP}. 
All the results in Table 2 have been obtained by  {\tt CompHEP} and cross checked
against our corrected codes.  

In Table 2 the cross sections are given for SUSY parameters and kinematic cuts as 
discussed above. In  column 2  we present the cross   section for 
$\ee \rightarrow \nu \bar{\nu} \gamma$. The first number corresponds to 
cross section of the bare process and the number in parenthesis to that 
corrected for initial state radiations. The latter is somewhat larger than the former
for reasons discussed above. The SUSY signal cross sections are presented in columns
3 and 4, while the total integrated luminosity 
at each energy as given by  Table 1, is presented in
column 5. It was already noted in Ref.\cite{asesh1} that the signal cross sections are
by and large unaffected by correcting for initial state radiation. 
The  number of background events without (with) initial state radiation
is given in column 6, while the total number of signal events (obtained from 
the information  in  column
3, 4 and 5) is given in column 7. It is quite clear that the signal($S$) to $\sqrt{{\rm 
background}(B)}$
ratio at any particular energy  is not statistically very significant. However,  
adding the  numbers in the last two columns
of Table 2 as shown in the last row,  
we find that $S / \sqrt{B}$ is 4.4 if  initial state radiation 
is taken into account. 

It may be recalled that the 
range of the sneutrino mass preferred by the precision electroweak data is 55 GeV $<
 m_{\snu} < $ 80 GeV. On the face of it, therefore, it appears 
as though  only the lower edge of this  mass range  can be probed.
However, there are some opportunities for improvement. First of all LEP
data above the $Z$-pole may be considered ( 
the present analysis is restricted to $\sqrt{s} > $ 189 GeV). 
In view of the approximate nature of  the present  analysis we made 
no attempt to optimize the cuts. For example, a fixed upper cut on $E_\gamma$ 
has been employed in the present analysis at all energies. Alternatively
cuts  suitably tailored for probing a particular 
  $m_{\snu}$ at a specific C.M.  energy may  be employed. Finally the contribution
of  $\ee \to \Chl \Chh  \gamma$, where $\Chh$ also decay  invisibly,  may further enhance
the signal.
 
There are  reasons to be optimistic about  this analysis. It is encouraging to
note that there is already an attempt\cite{ask} to combine the single 
photon + missing energy data, at all energies above the $Z$-pole, 
of  the four LEP groups to constrain some new physics scenarios.
From the combined recoil mass (M) distribution it even appears that there is  a modest
excess of events over the SM background for M$>$ 160 GeV. 
It should be relatively straightforward to extend this analysis to the case of a 
pair of heavy invisible sneutrinos produced in association with the photon.     
  If no signal is found, some sneutrino masses in the range preferred by the precision data
will be excluded. Even this is an  improvement over  the current  weak lower bound 
on $m_{\snu}$ ($\approx 0.5 m_Z$) obtained from the  $Z$-pole data.
Of course the latter bound is model independent, while one under consideration
depends on gaugino mass unification and on assumptions about $\mr$. 
Yet for the first time one will 
probe $m_{\snu}$  through direct  production of invisible sneutrino pairs even if sleptons
are beyond  the kinematic reach of LEP. 
In that sense the model dependent assumptions employed are quite different from 
the assumptions implicit in mSUGRA. The bounds obtained by this method will therefore
be complimentary to that obtained, e.g., in Ref.\cite{aleph}. How a completely model
independent bound can be obtained by this method will be discussed below.

Since $\mr$ was somewhat arbitrarily chosen in our  analysis, some discussion 
of the sensitivity of the signal to  this parameter is in order. This parameter
does not affect the cross section of $\ee \to \snu \snu^*  \gamma$ which is by far 
the dominant 
contribution to the signal for low and moderate $m_{\snu}$. 
However, if  $m_{\snu}$ is close to the upper
end of the range preferred by the electroweak precision data,  
the observability of the signal can indeed be  affected by $\mr$ as can be seen
from Figure 1 
 drawn for $\sqrt{s}$ = 205 GeV. The upper (lower) curve corresponds to $\mr = \ml$
($\mr = 1.5 \ml$). The other SUSY parameters are as in Table 2. 
The background cross section  at this energy is given in  Table 2.

\vskip 15pt
\begin{figure}[htbp]
\begin{center}
\vskip-5.5cm
\mbox{\centerline{\epsfig{file=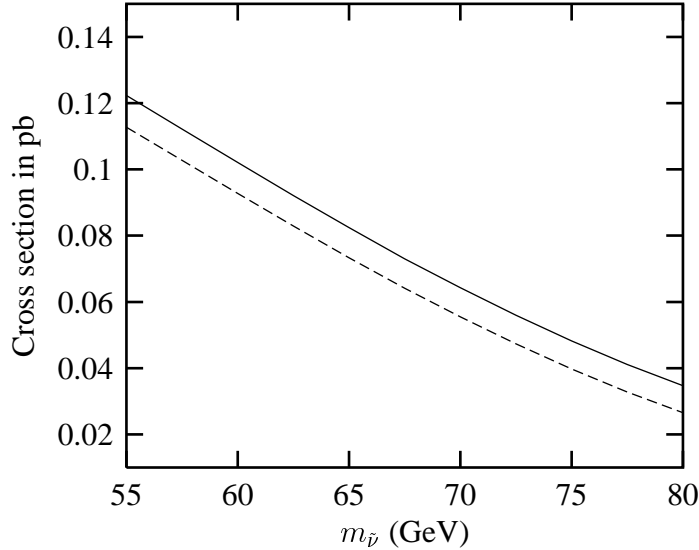,width=21cm}}}
\end{center}
\vspace*{-18cm}
\caption{Variation of the signal cross section with $m_{\snu}$ (see text for further
details)}
\end{figure}

One can  use the absence of  signal to put a lower bound on the mass of any 
invisible sneutrino   in a model independent way. 
 If we do not make any  additional assumption like gaugino mass unification
 then the neutralino contributions  to the $\gamma
 + $ missing energy signal is  uncorrelated to the sneutrino contribution.
The most conservative lower bound on the sneutrino mass can, therefore, be obtained by
comparing the  data with the sneutrino contribution
alone, which is independent of $\mr$.

For $\snu_{\mu}$ and $\snu_{\tau}$ only the $Z$-mediated $s$-channel 
diagrams contribute. Hence, the cross sections are model independent
and depend on $m_{\snu}$ alone. 
The $\snu_{e}$ 
 contribution  depends on $m_{\snu}$, $M_2$, $\mu$ and tan $\beta$.
The last three
parameters determine the chargino mass and its coupling with the sneutrino.
The amplitude has two pieces: the $s$-channel  $Z$-exchange terms
and the $t$-channel chargino exchange terms. The key observation is that
the interference among  the
$s$-channel and $t$-channel diagrams  are always constructive for all choices of
$M_2$, $\mu$ and tan $\beta$. Thus the smallest possible signal cross section 
can be obtained by neglecting the contribution of the $t$-channel diagrams. This
corresponds to infinite chargino mass or Higgsino dominated charginos, completely
decoupled from the sneutrinos. Now one obtains the minimum  signal as a 
function of $m_{\snu}$ only by assuming three generations of degenerate sneutrinos.

Considering  only the $s$-channel contribution to the signal we find 
96(64)
events for $m_{\snu}$ = 50(55) GeV for the entire energy range shown in Table 2.
Such a small signal will be swamped by the background. Obviously the current 
data alone is not adequate  to significantly strengthen the existing 
model independent lower bound on $m_{\snu}$. However, the size of the signal at
the Next Linear Collider as given in \cite{asesh1} suggests that if no signal is found
a model independent bound may  emerge.
The inclusion  of other  SUSY contributions  in specific  models, 
can only strengthen this bound. 

 Light sneutrinos in the mass range indicated by the precision electroweak
data may dramatically influence the SUSY search strategies 
at the upgraded Tevatron.
It has been already been noted  since long back,
 that the hadronically quiet trilepton + $\ET$ event 
resulting from $\Chl \Chh$ pairs is the best channel for SUSY search at this
machine\cite{tev}. In the presence of light sneutrinos, however, many of
the $\Chh$s will decay into the invisible channel $\nu - \snu^*$. Thus the
trilepton channel may be considerablly weakened or wiped out depending on
the BR of the invisible channel, which is model dependent. On the other
hand $\wino$ pair also has a healthy production cross section at the upgraded Tevatron
\cite{tev}. Since these charginos will decay into the two body channel $l + \snu$
with a large BR, the opposite sign dilepton + $\ET$ signal will be enhanced
\cite{surajit}.
Suitable kinematical cuts can show this signal over WW and other relevant
backgrounds. From the results of \cite{surajit} it appears that the entire
range of $m_{\snu}$ preferred by the elctroweak data can be probed.

\vskip 5pt
\noindent
{\bf Acknowledgements :} AD thanks Professor H. Dreiner and other members of the
Theoretical Physics Group, University of Bonn where part of the work was done.
He thanks Professor M. Kobel for discussions.  
Aseshkrishna Datta is supported by French
MNERT fellowship and subsequently by CNRS France. He also likes to thank
the Theoretical Physics Group of Abdus Salam ICTP, Trieste for a visit during summer 
2001 where this project was started.


\begin{thebibliography}{99}
\bibitem{lumi} The total luminosity collected by each of the LEP groups 
in  a range of c.m. energies  can
be found from their publications. However, the break-ups for each c.m energy are not always
readily available (see Table 1 and the discussions preceeding it). Some of the useful references 
are DELPHI Collaboration in\cite{photon};
; ALEPH collaboration in\cite{photon}.
\end{thebibliography}
\end{document}